\documentclass[%
 reprint,
superscriptaddress,
 amsmath,amssymb,
 aps,
]{revtex4-2}

\usepackage{graphicx}% Include figure files
\usepackage{dcolumn}% Align table columns on decimal point
\usepackage{bm}% bold math
\usepackage{hyperref}% add hypertext capabilities
\newcommand{\etal}{{\em et al.}}
\newcommand{\pbo}  {PbWO$_4$}

\begin{document}
\title{A new direct detection electron scattering experiment to search for the X17 particle.}
\author{D.~Dutta}	\affiliation{Mississippi State University}
\author{H.~Gao}	\affiliation{Duke University}
\author{A.~Gasparian}	\affiliation{North Carolina A\&T State University}
\author{T. J. Hague}    \affiliation{North Carolina A\&T State University} \affiliation{Lawrence Berkeley National Laboratory}
\author{N.~Liyanage}	\affiliation{University of Virginia}
\author{R.~Paremuzyan}	\affiliation{Thomas Jefferson National Accelerator Facility}
\author{C.~Peng}	\affiliation{Argonne National Laboratory}
\author{W.~Xiong}	\affiliation{Shandong University}
\author{P.~Achenbach} \affiliation{Thomas Jefferson National Accelerator Facility}
\author{A. Ahmidouch}	\affiliation{North Carolina A\&T State University}
\author{S. Ali}	\affiliation{University of Virginia}
\author{H. Avakian}	\affiliation{Thomas Jefferson National Accelerator Facility}
\author{C. Ayerbe-Gayoso}	\affiliation{Mississippi State University}
\author{X. Bai}	\affiliation{University of Virginia}
\author{M. Battaglieri}	\affiliation{Isituto Nazionale di Fisica Nucleare, Sezione di Genova}
\author{H. Bhatt}	\affiliation{Mississippi State University}
\author{A. Bianconi}	\affiliation{Dipartimento di Ingegneria dell'Informazione, Università di Brescia}   \affiliation{Isituto Nazionale di Fisica Nucleare, sezione di Pavia}
\author{J. Boyd}	\affiliation{University of Virginia}
\author{D. Byer}	\affiliation{Duke University}
\author{P. L. Cole}	\affiliation{Lamar University}
\author{G. Costantini}	\affiliation{Dipartimento di Ingegneria dell'Informazione, Università di Brescia}	\affiliation{Isituto Nazionale di Fisica Nucleare, sezione di Pavia}
\author{S. Davis}	\affiliation{North Carolina A\&T State University}
\author{M. De Napoli}	\affiliation{Isituto Nazionale di Fisica Nucleare, Sezione di Catania}
\author{R. De Vita}	\affiliation{Isituto Nazionale di Fisica Nucleare, Sezione di Genova}
\author{B. Devkota}	\affiliation{Mississippi State University}
\author{B. Dharmasena}	\affiliation{University of Virginia}
\author{J. Dunne}	\affiliation{Mississippi State University}
\author{L. El Fassi}	\affiliation{Mississippi State University}
\author{V. Gamage}	\affiliation{University of Virginia}
\author{L. Gan} \affiliation{University of North Carolina, Wilmington}
\author{K. Gnanvo}	\affiliation{Thomas Jefferson National Accelerator Facility}
\author{G. Gosta}	\affiliation{Dipartimento di Ingegneria dell'Informazione, Università di Brescia}	\affiliation{Isituto Nazionale di Fisica Nucleare, sezione di Pavia}
\author{D. Higinbotham}	\affiliation{Thomas Jefferson National Accelerator Facility}
\author{C. Howell}	\affiliation{Duke University}
\author{S. Jeffas}	\affiliation{University of Virginia}
\author{S. Jian}	\affiliation{University of Virginia}
\author{A. Karki}	\affiliation{Mississippi State University}
\author{B. Karki}	\affiliation{Duke University}
\author{V. Khachatryan}	\affiliation{Duke University}
\author{M. Khandaker}	\affiliation{Energy Systems}
\author{V. Kubarovsky}	\affiliation{Thomas Jefferson National Accelerator Facility}
\author{I. Larin}	\affiliation{University of Massachusetts}
\author{M. Leali}	\affiliation{Dipartimento di Ingegneria dell'Informazione, Università di Brescia}	\affiliation{Isituto Nazionale di Fisica Nucleare, sezione di Pavia}
\author{V. Mascagna}	\affiliation{DiSAT, Università dell'Insubria}	\affiliation{Isituto Nazionale di Fisica Nucleare, sezione di Pavia}
\author{G. Matousek}	\affiliation{Duke University}
\author{S. Migliorati}	\affiliation{Dipartimento di Ingegneria dell'Informazione, Università di Brescia}	\affiliation{Isituto Nazionale di Fisica Nucleare, sezione di Pavia}
\author{R. Miskimen}	\affiliation{University of Massachusetts}
\author{P. Mohanmurthy}	\affiliation{Mississippi State University}
\author{H. Nguyen}	\affiliation{University of Virginia}
\author{E. Pasyuk}	\affiliation{Thomas Jefferson National Accelerator Facility}
\author{A. Rathnayake}	\affiliation{University of Virginia}
\author{J. Rittenhouse West}	\affiliation{Lawrence Berkeley National Laboratory}
\author{A. Shahinyan}	\affiliation{Yerevan Physics Institute}
\author{A. Smith}	\affiliation{Duke University}
\author{S. Stepanyan}	\affiliation{Thomas Jefferson National Accelerator Facility}
\author{E. van Nieuwenhuizen}	\affiliation{Duke University}
\author{L. Venturelli}	\affiliation{Dipartimento di Ingegneria dell'Informazione, Università di Brescia}	\affiliation{Isituto Nazionale di Fisica Nucleare, sezione di Pavia}
\author{B. Yu}	\affiliation{Duke University}
\author{Z. Zhao}	\affiliation{Duke University}
\author{J. Zhou}	\affiliation{Duke University}

\affiliation{Lawrence Berkeley National Laboratory}
\date{January 2023}

\begin{abstract}
A new electron scattering experiment (E12-21-003)~\cite{Ahmidouch:2021edo} to verify and understand the nature of hidden sector particles, with particular emphasis on the so-called X17 particle, has been approved at Jefferson Lab. 
The search for these particles is motivated by new hidden sector models introduced to account for a variety of experimental and observational puzzles: excess in $e^+e^-$ pairs observed in multiple nuclear transitions, the 4.2$\sigma$ disagreement between experiments and the standard model prediction for the muon anomalous magnetic moment, and the small-scale structure puzzle in cosmological simulations.
The aforementioned X17 particle has been hypothesized to account for the excess in $e^+e^-$ pairs observed from the $^8$Be M1, $^4$He M0, and, most recently, $^{12}$C E1 nuclear transitions to their ground states observed by the ATOMKI group. 
This experiment will use a high resolution electromagnetic calorimeter to search for or set new limits on the production rate of the X17 and other hidden sector particles in the $3 - 60$ MeV mass range via their $e^+e^-$ decay (or $\gamma\gamma$ decay with limited tracking). 
In these models, the $1 - 100$ MeV mass range is particularly well-motivated and the lower part of this range still remains unexplored. 
Our proposed direct detection experiment will use a magnetic-spectrometer-free setup (the PRad apparatus) to detect all three final state particles in the visible decay of a hidden sector particle allowing for an effective control of the background and will cover the proposed mass range in a single setting. 
The use of the well-demonstrated PRad setup allows for an essentially ready-to-run and uniquely cost-effective search for hidden sector particles in the $3 - 60$ MeV mass range with a sensitivity of 8.9$\times$10$^{-8}$ - 5.8$\times$10$^{-9}$  to $\epsilon^2$, the square of the kinetic mixing interaction constant between hidden and visible sectors.

\end{abstract}

\maketitle
\section{Introduction}
% The remarkable fact that $\sim$~85\% of the matter in the Universe is of unknown origin - termed dark matter (DM) - is inferred from astronomical measurements over a wide range of distance and time scales, from the Milky Way to the largest cosmological structures and as early as Big Bang Nucleosynthesis to today. Yet all of the evidence thus far is gravitational, providing no direct information about the identity of dark matter. Consequently, the investigation into the nature of DM, from its origin to its composition and how it interacts with other forces both old and new, is one of the grand challenges in fundamental science. There are many candidate theories for dark matter and dark sector mediators that span an enormous mass range, from $10^{-22}$ eV to at least 100 times the mass of the sun. 
% Several recent observations and anomalies have required new dark matter interactions and candidates, such as hidden sector dark matter (HSDM)~\cite{Alexander:2016hsr} models, that point to the $1 - 100$ MeV/c$^2$ mass range as well-motivated for high priority searches~\cite{Battaglieri:2017aum}. 
% In this proposal, we describe an experiment that will search in the 3 - 60 MeV mass region for heretofore unobserved hidden sector particles. 
% This experiment will utilize the bremsstrahlung-like production of a hidden sector boson that subsequently decays to a $e^+e^-$ or $\gamma\gamma$ pair (i.e. visible decays) which will be detected in the PRad setup with minimal modifications.

On large distance scales, the structure of the Universe as inferred from cosmological data is consistent with dark matter particles that are cold, collision-less, and interact with ordinary matter purely via gravity~\cite{Tulin:2017sss}. Decades of cosmological data have converged on the standard model of cosmology, dubbed $\Lambda \rm CDM$, with cold dark matter (CDM) as a crucial ingredient ~\cite{Bachall:99cdm} along with dark energy (the cosmological constant of general relativity, $\Lambda$) and ordinary matter. Weakly interacting massive particles (WIMPs) have long been a primary candidate to explain dark matter (DM)~\cite{jungman96,Feng:2010gw}. However, recent searches have ruled out a large segment of the phase space for WIMPs~\cite{Battaglieri:2017aum,Aprile:2018x1t} leading to new theoretical developments~\cite{Alexander:2016hsr}. 
Additionally, several anomalous results in recent years have pointed to holes in our understanding of the nature of matter, discussed further in Sec.~\ref{sec:motivation}.

These null results and the anomalous results have led to an increased interest in dark mediator models.  In typical models these mediator particles ($X$) interact with the standard model (SM) through  a `kinetic mixing' mechanism creating a portal between the hidden sector and the SM. In most models the $X$ particle couples to the electric charges of SM matter. Some newer models propose that the coupling to SM is tied to flavor of quarks or leptons and not via the electric charge. A common thread between many of the observed anomalies is that they can be explained by a new MeV-scale dark mediator particle. Given the convergence of such models on the $1-100$ MeV mass range for a new particle, it is critical to focus experimental searches to this mass range in order to resolve the anomalies.

The availability of a high duty factor, high luminosity electron beam at Jefferson Lab provides an ideal setup to search for MeV-scale dark mediators with small coupling constants. 
The well tested PRad setup in Hall B~\cite{PRad:2019} will be used in this experiment to reach our physics goals.
Using a magnetic-spectrometer-free setup allows the experiment to be sensitive to the full mass range in a single experimental setting. The detection of all three final state particles in the \pbo{} calorimeter along with tracking with GEM chambers allows for an effective control of the backgrounds. Moreover, it provides an essentially ready-to-run and uniquely cost effective search for hidden-sector particles in the 3 - 60 MeV mass range.

\section{Physics Motivation}
\label{sec:motivation}
%Not copy/pasting yet, this should be heavily edited down

% \textbf{PLACEHOLDER: X17 is most important thing here. 3 measurements saw it. talk about that}

% A 1996 experiment~\cite{deBoer1996}, using the 2.5 MV Van de Graaff accelerator at Institut f{\"u}r Kernphysik of the University of Frankfurt, noted a $4.5\sigma$ deviation from expectation in the angular distribution of $e^+e^-$ from Internal Pair Conversion (IPC) of the $^8$Be M1 resonance.
% Analysis and Simulations of the signal that was seen led to the conclusion that a neutral boson of mass 9 MeV was a possible explanation that could not be ruled out by existing constraints~\cite{deBoer1997,deBoer:2001sjo}.

The primary motivation for this new experiment is to resolve the X17 anomaly. 
A 2015 experiment~\cite{Krasznahorkay:2015iga} at the ATOMKI 5 MV Van de Graaff accelerator sought  to study a previous 9 MeV anomaly~\cite{deBoer1996,deBoer1997,deBoer:2001sjo}.
While this experiment nullified and explained the previous result, it saw an excess of $e^+e^-$ pairs beyond the expectation of internal pair creation (IPC) for a different mass.
A subsequent analysis of these results has shown that the $6.8\sigma$ anomaly is consistent with a new particle with a mass of 16.84 MeV, dubbed X17. 
A followup experiment by the ATOMKI group was conducted on the 20.01 MeV $0^-\to0^+$ transition in $^4$He.
The results for $^4$He also reports an $e^+e^-$ excess consistent with the so-called X17 particle~\cite{Krasznahorkay:2019lyl,Krasznahorkay:2021joi}.
In a new 2022 experiment, the group has reported the anomaly in the $^{12}$C E1 transition~\cite{Krasznahorkay:2022pxs}.
Of note, this most recent result agrees with the previous masses but disagrees with the measured branching ratios.

\begin{figure}[hbt!]
    \centering
    \includegraphics[width=\columnwidth]{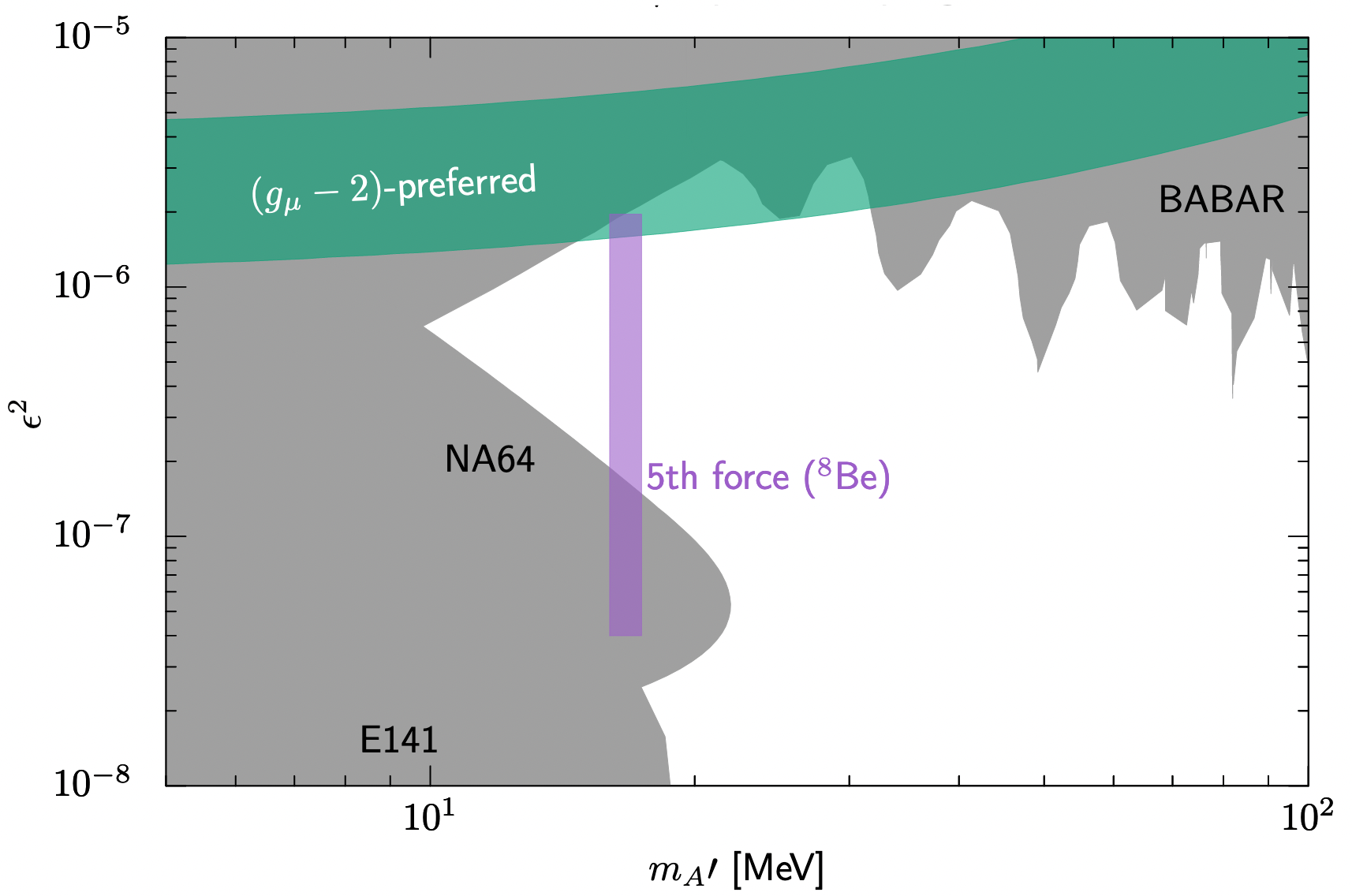}
    \caption{Current constraints on a fifth force explanation of the $^8$Be anomaly.  The vertical axis is the leptonic coupling strength relative to $\alpha_{QED}$, with horizontal axis the mass of the mediator. Excluded regions, in gray, are taken from measurements that depend solely on leptonic interactions. Dark photon exclusions via hadronic measurements are not shown.
    Reproduced from Ref.~\cite{darklight:2020pp}.}
    \label{fig:be_anom}
\end{figure}

Feng \etal~\cite{Feng:2016jff} analyzed the X17 signal against existing constraints.
The proposed explanation is that the signal is from the decay of a protophobic gauge boson that mediates a fifth force with a length scale of 12~fm.
This explanation can also possibly explain the muon anomalous magnetic moment and an excess of $\pi^0\to e^+e^-$ decays~\cite{KTeV:2006pwx}. The current constraints from leptonic production mechanisms, where the effective coupling to a new force-carrier is proportional to electric charge, are shown in Fig.~\ref{fig:be_anom}. For more generic fifth-force models with  quark flavor-dependent couplings~\cite{Feng:2016jff}, a much wider parameter space with multiple couplings must be considered.

The fifth force based explanations are being challenged by recent reanalyses~\cite{Zhang:2020ukq,Kubarovsky:2022zxm} and the observed discrepancies could be the result of as-yet-unidentified nuclear reactions, excited diquark states, or experimental effects. Nonetheless, these results have garnered a lot of attention, and must be independently validated with the highest urgency.

% \textbf{PLACEHOLDER: muon g-2. loop diagrams. Small paramater-space area near 17 MeV could explain both.}

The Muon $g-2$ collaboration has recently reported their measurement of the muon magnetic moment at Fermi National Accelerator Laboratory (FNAL) which is consistent with their previous measurement at Brookhaven National Laboratory (BNL)~\cite{Albahri:2021kmg,Bennett:2006fi]}.
These two results, with their uncertainties combined, show a $4.2\sigma$ deviation from the Standard Model prediction.

There have been several proposals as to new physics that could lead to this deviation.
One such class of explanations shows that undiscovered hidden sector particles can directly contribute to loop corrections in the calculation of $(g-2)_{\mu}$. 
The contributions of a hidden sector particle to the $(g-2)_{\mu}$ scales as $1/m_{X}^2$, thus a smaller mass will have a comparably larger contribution than a higher mass particle. 
There have been numerous publications proposing new particles (such as dark photons) in the MeV-range that, at a minimum, partially account for the muon anomalous magnetic moment~\cite{Arcadi:2021yyr,Athron:2021iuf,Borah:2021jzu,Ge:2021cjz}.
Fig.~\ref{fig:be_anom} shows that there is an overlap in the phase space for a particle that would simultaneously resolve the X17 and $(g-2)_{\mu}$ anomalies.

% \textbf{PLACEHOLDER: Streamline stuff below. DM is interesting and a motivator but we don't want to sell this as DM experiment.}

Several recent models have suggested that the coupling of the X17 to SM is tied to flavor of quarks or leptons and not via the electric charge.
One such example is the protophobic X17 proposed in Ref.~\cite{Feng:2016jff} used to explain the X17 anomaly.
If an X17 were to have flavor-dependent couplings the parameter space for it is more open than previously thought \cite{Feng:2022inv}. These new possibilities point to an urgent need for independent verification of the X17 anomaly.  

\section{Experimental Method and Setup}

\begin{figure}[hbt!]
    \centering
    \includegraphics[width=0.35\columnwidth]{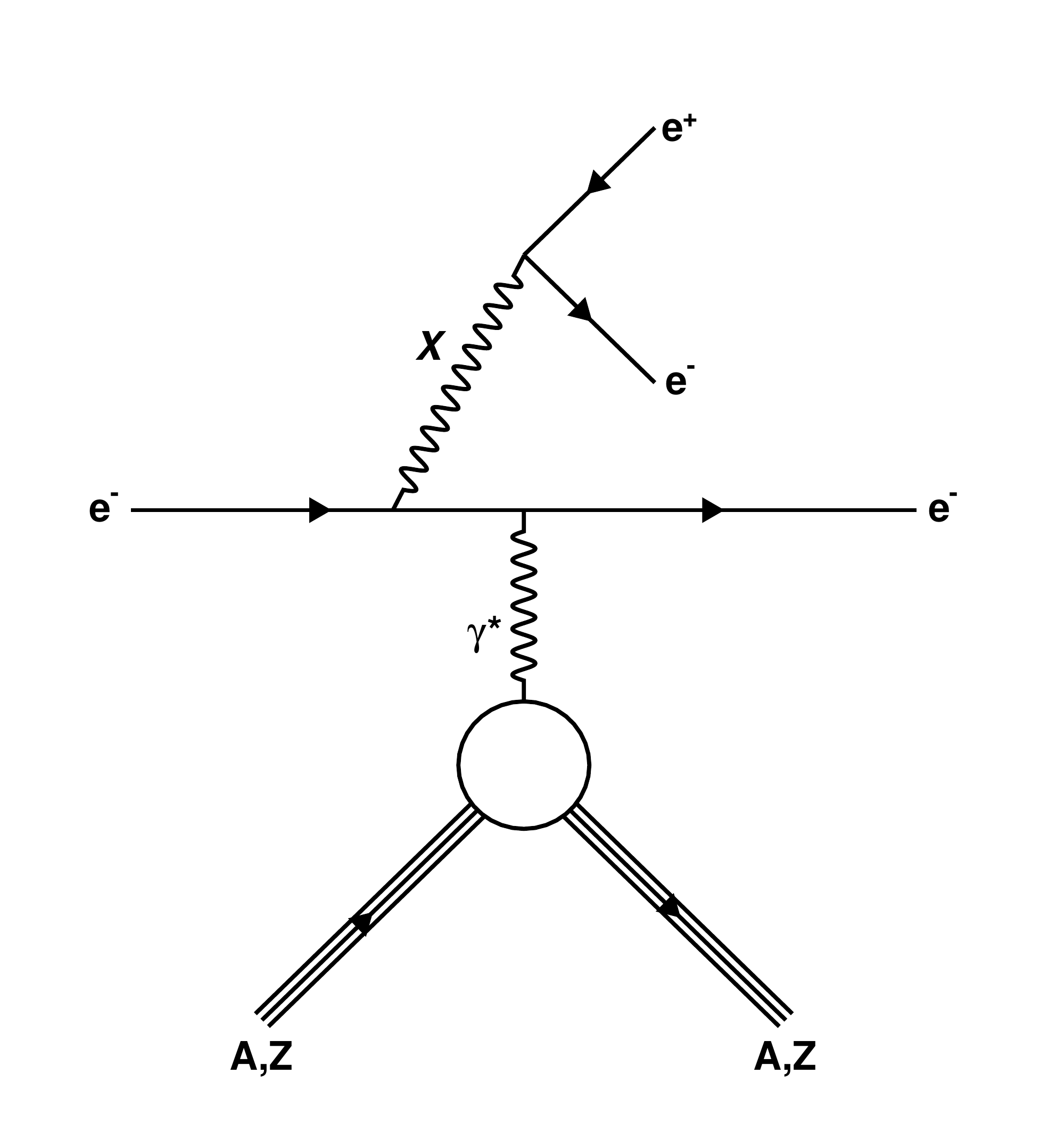}
    \includegraphics[width=0.35\columnwidth]{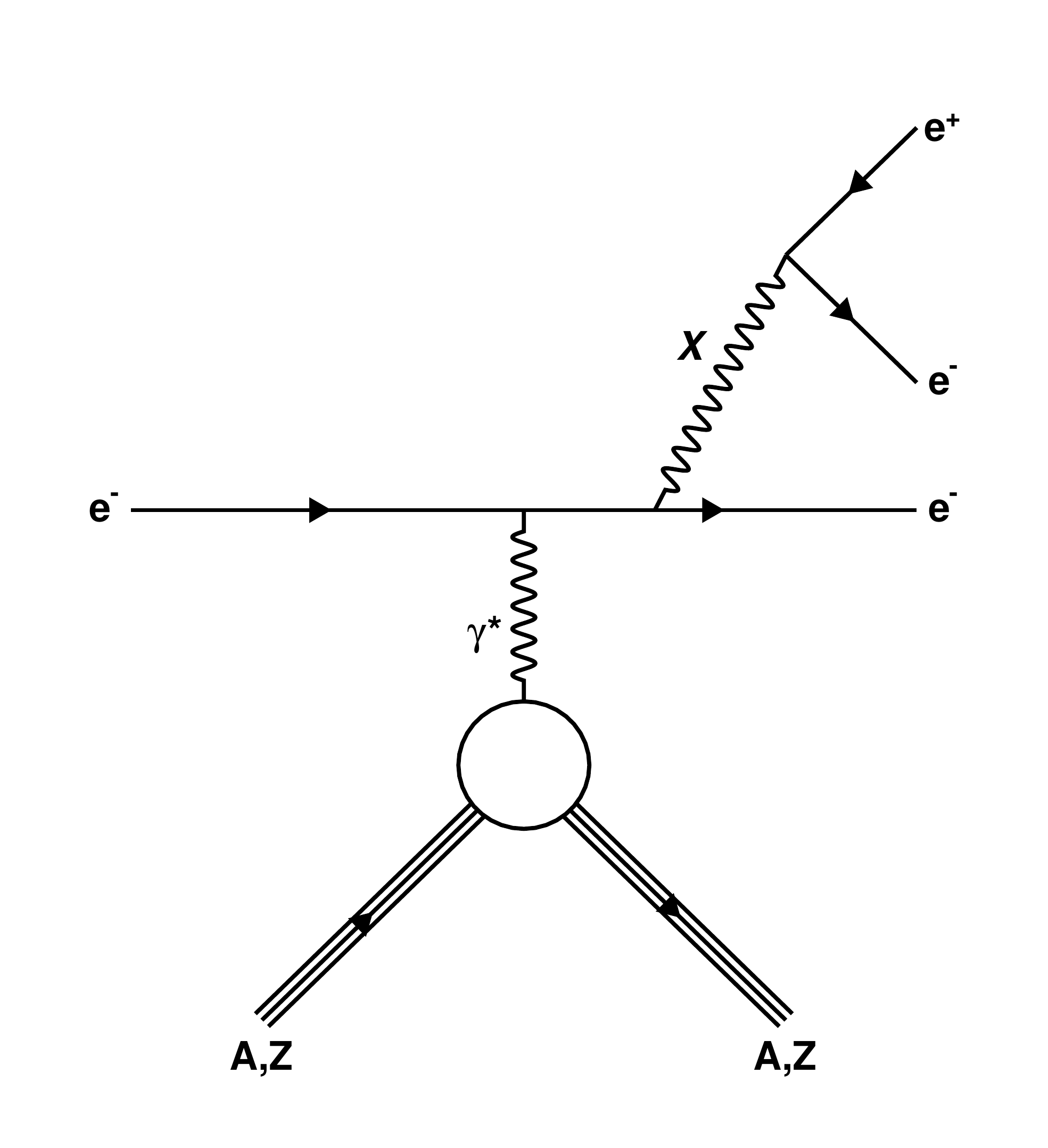}
    \caption{Bremsstrahlung-like production of a hidden sector force carrier $X$ from electron scattering. The left diagram shows production from the incoming electron and the right diagram shows production from the scattered electron.}
    \label{fig:bremsstrahlung-feynman}
    %will be filled in later
\end{figure}

\begin{figure}[hbt!]
    \centering
    \includegraphics[width=0.3\columnwidth]{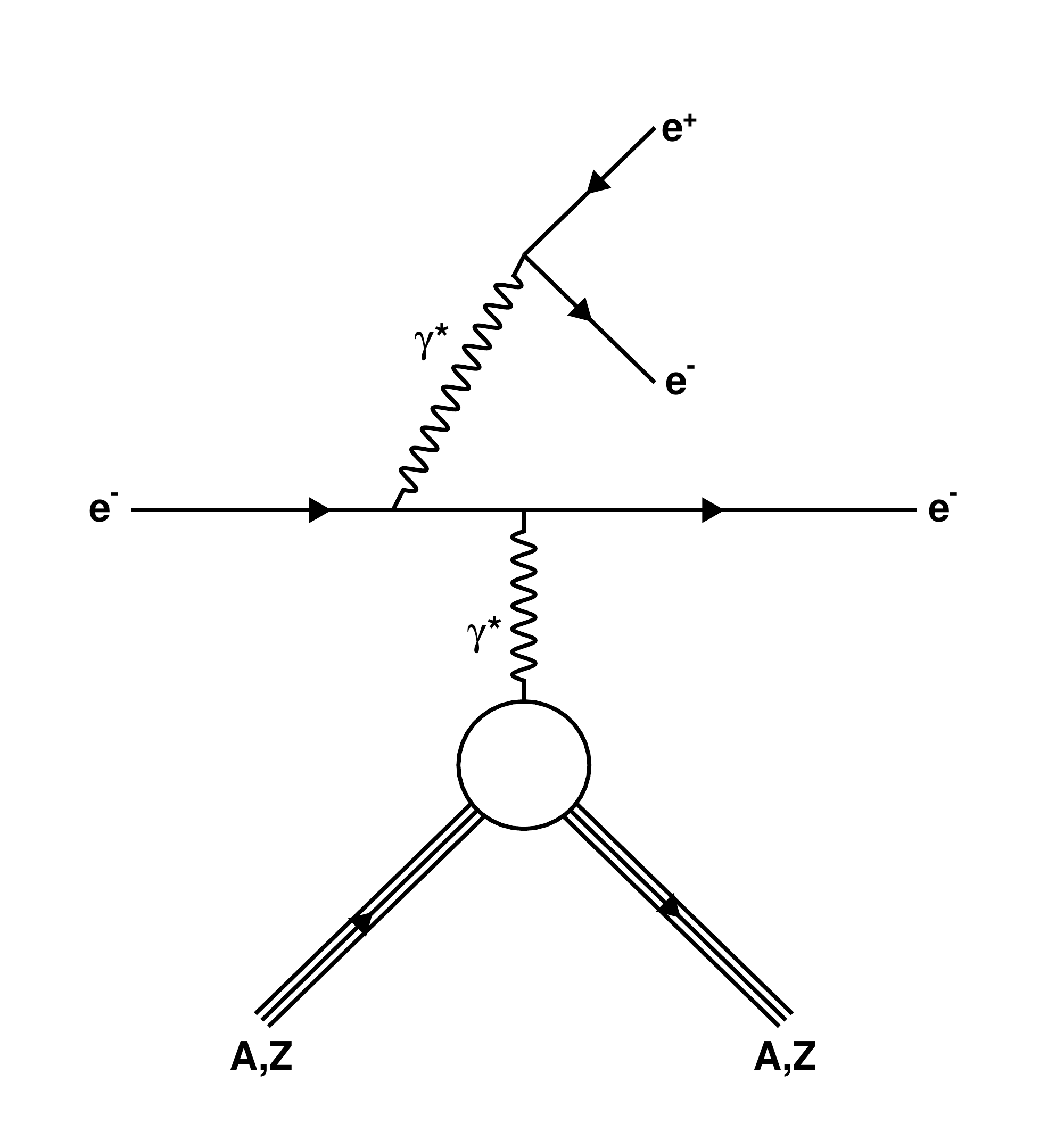}    \includegraphics[width=0.3\columnwidth]{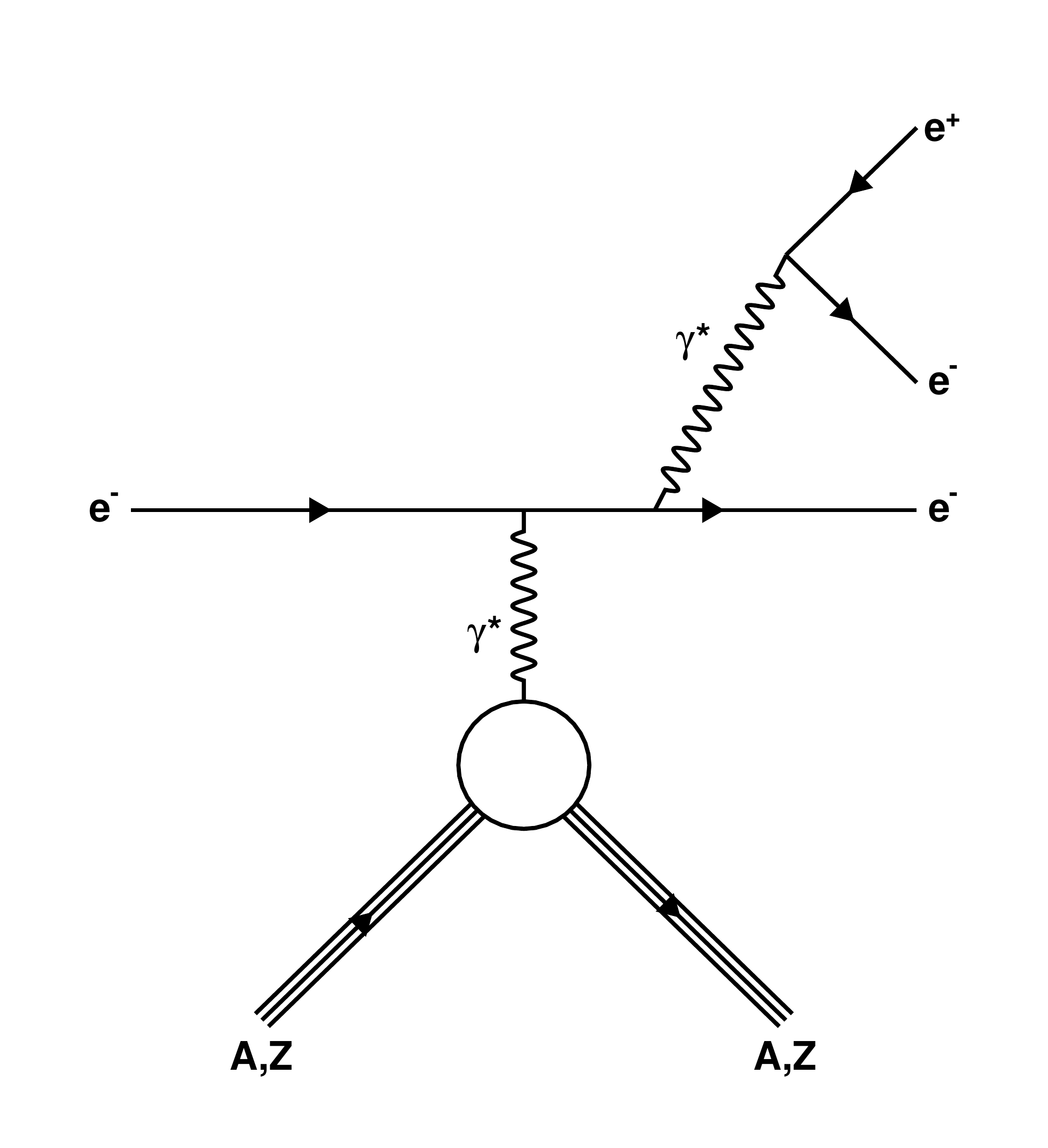}
    \includegraphics[width=0.3\columnwidth]{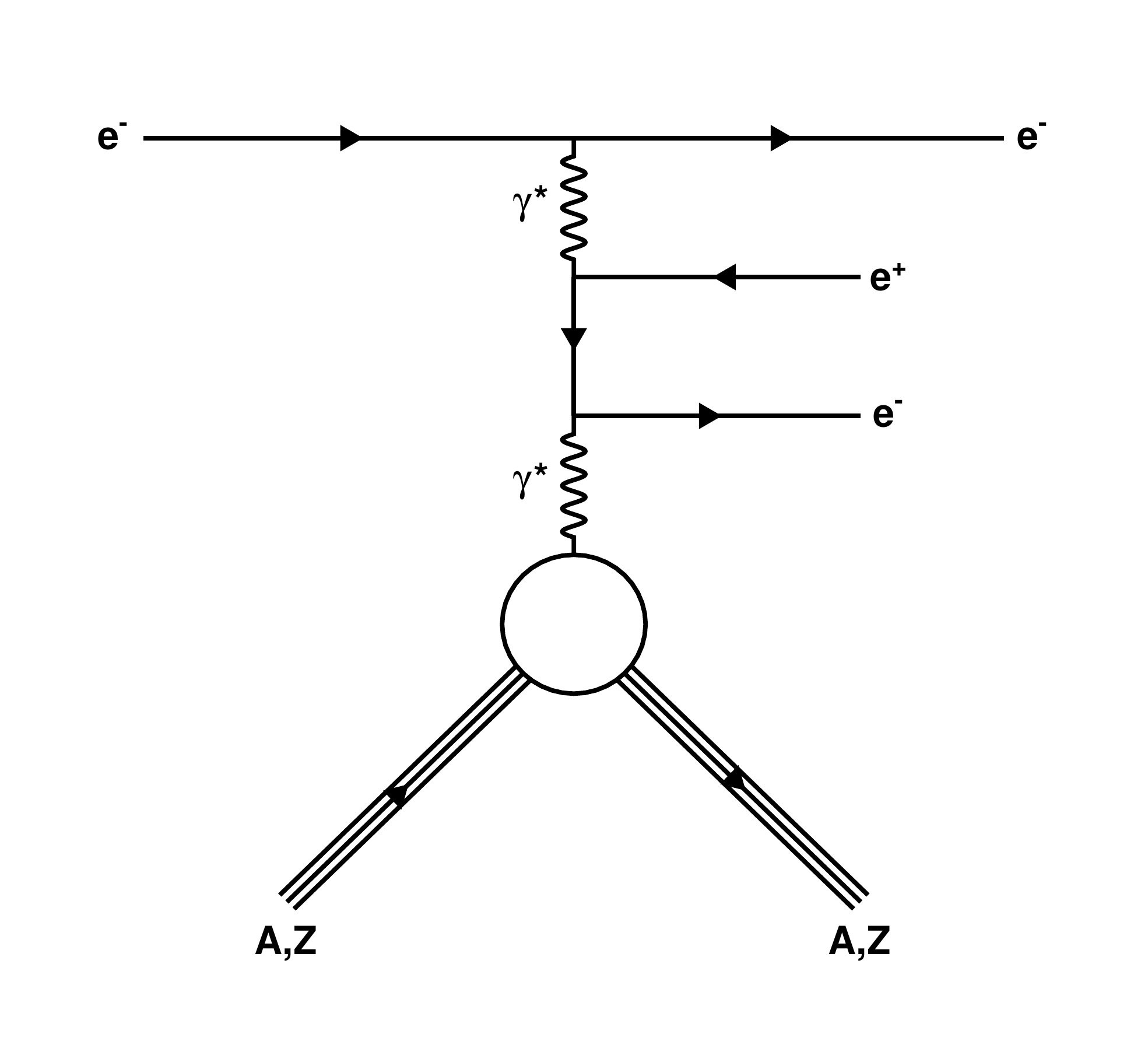}
    \caption{The QED background from radiative (left and middle) and Bethe-Heitler (right) process.}
    \label{fig:background-feynman}
    %will be filled in later
\end{figure}

This experiment will focus on the bremsstrahlung-like production of hidden sector particles from the initial electron or the scattered electron (both shown in Fig.\ref{fig:bremsstrahlung-feynman}), in the $3-60$ MeV mass range.
When searching for new particles, it is of the utmost importance to minimize backgrounds in order to prevent false ``bumps'' in the mass spectrum. The primary QED background in this experiment (see Fig.~\ref{fig:background-feynman}) are from the radiative pair production which is an irreducible background and the Bethe-Heitler trident reactions which can be kinematically suppressed.

\begin{figure}[!hbt]
%\vskip 0.35truecm
\centerline{
\includegraphics[width=\columnwidth]{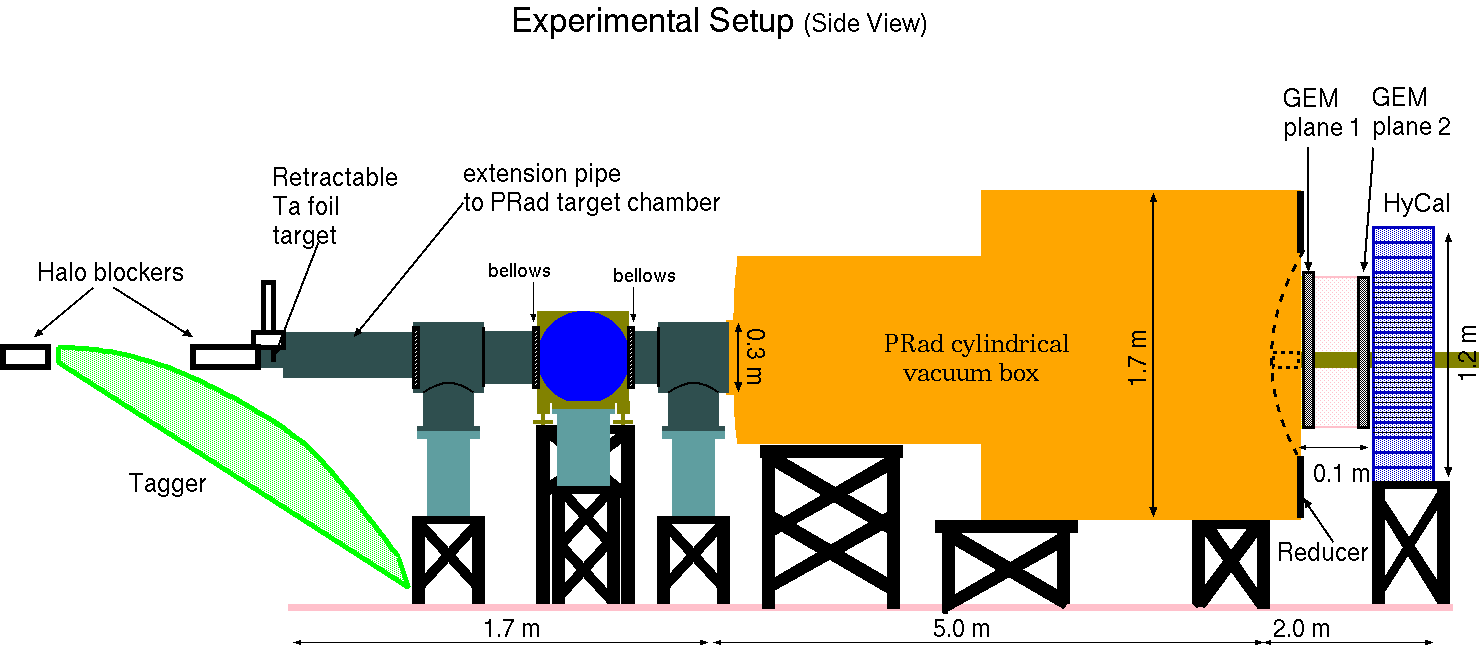}}
\caption{Schematic of the experimental setup.}
\label{fig:expt_setup}
\end{figure}

The experiment plans to reuse the PRad setup (shown in Fig.~\ref{fig:expt_setup}), with the addition of a Tantalum foil target placed 7.5 m upstream of the calorimeter. The high resolution \pbo ~crystal part of the HyCal electromagnetic calorimeter~\cite{Gasparian:2002px,Gasparian:2004xa} will be used together with a new flash ADC (FADC) based readout system for the calorimeter. %A thin scintillator detector placed ~0.9~m from the target will help tag the charged particles originating from the target. 
A 2.2 and 3.3 GeV Continuous Wave (CW) electron beam from CEBAF will be incident on a retractable ultra-thin target consisting of a 1~$\mu$m Tantalum foil. The scattered particles will traverse the 7.5 m long flight path in vacuum, en route to a pair of common ionization volume GEM chambers coupled to the HyCal EM calorimeter. The vacuum chamber will consist of the PRad target chamber (or an appropriate diameter beam pipe) and the 5~m long PRad vacuum chamber with a thin window.
%Just as in the PRad experiment, the scattered electrons will travel through the 5~m long vacuum chamber  to minimize multiple scattering and backgrounds. 
The vacuum chamber matches the geometrical acceptance of the calorimeter. 
An extension piece will be added to the upstream end to couple the PRad target chamber to the super-harp which will now hold the target foil. A reducer ring will be attached to the downstream exit of the PRad vacuum chamber and a new 1~mm thick Al exit window will be used such that it matches the \pbo ~portion of the calorimeter. Two layers of GEM detectors  will add a modest tracking capability to help reduce the photon background and to reduce the background originating from the vacuum chamber exit window.

The experiment will use a ``bump hunt'' technique in the direct detection search of heretofore unknown MeV mass particles. 
%The experiment is designed based on the kinematic constraints described above.  %A thin (2.5 mm) scintillator detector will be placed inside the PRad target chamber to tag all events originating from the target. 
All 3 cluster events with individual cluster energy within (0.02-0.85)$\times E_{beam}$ and with the sum of total energy deposited $E_{sum} >$ 0.7$\times E_{beam}$ will be recorded and examined for ``bumps'' in the  $M_{e^{+} e^{-}}$ (or $M_{\gamma \gamma}$) 
 invariant mass spectrum reconstructed from these events. This will allow for the $X$-particles production by virtual photons over a wide energy range in the forward solid angle coverage of the \pbo{} calorimeter. The capability of detecting events produced in a wide energy and angle range, in a single experimental setting, is one of the important features of our experiment which stands in contrast to all other magnetic spectrometer methods. The tracking provided by the pair of GEM chambers 
%along with the tag from the scintillator detector 
will be used to suppress background events from the experimental setup, particularly, the large area window at the exit of the vacuum chamber. The GEM chambers will also be used to veto any neutral particles. Only the \pbo{} part of the HyCal calorimeter will be used in the experiment to ensure high invariant mass resolution. %The lower resolution of the lead-glass part makes it unusable in this experiment. 
The experimental method discussed here applies directly to any spin-1 boson in the hidden sector that decay directly to a lepton pair. 

\section{Projected Results}

\label{sec:projected_results}
\begin{figure}[hbt!]
  \includegraphics[width=\columnwidth]{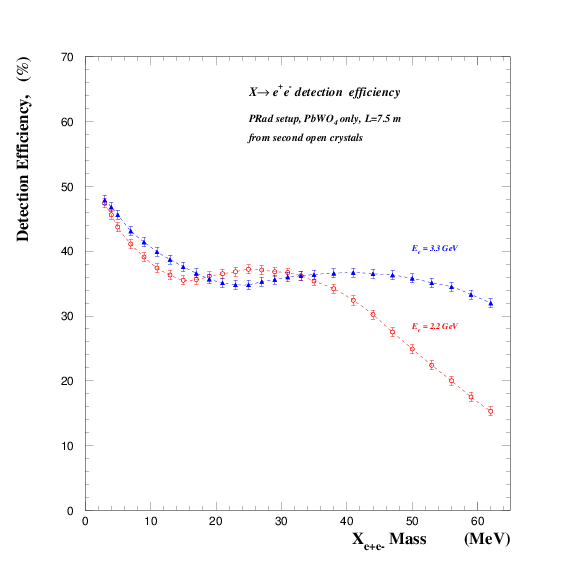}
  \caption{Acceptance of the spectrometer setup for both 2.2 and 3.3 GeV beam energies.}
  \label{fig:deteff}
\end{figure}

\begin{figure}[hbt!]
  \includegraphics[width=\columnwidth]{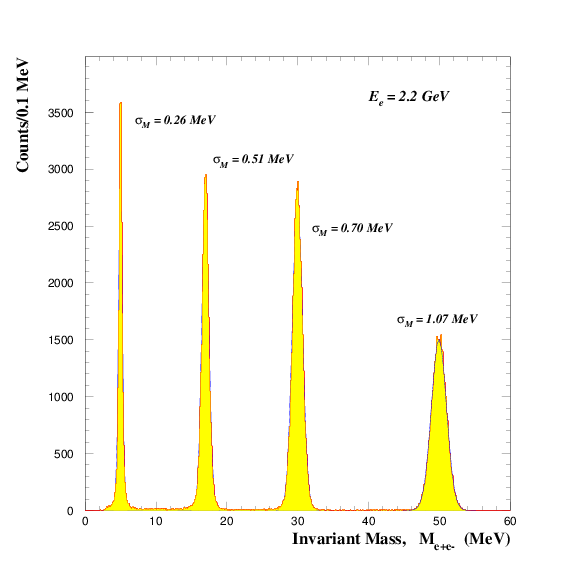}
  \caption{Invariant mass resolution of the spectrometer for 2.2 GeV beam energy.}
  \label{fig:invmres}
\end{figure}
\begin{figure}[b!]
  \centerline{
  \includegraphics[width=\columnwidth]{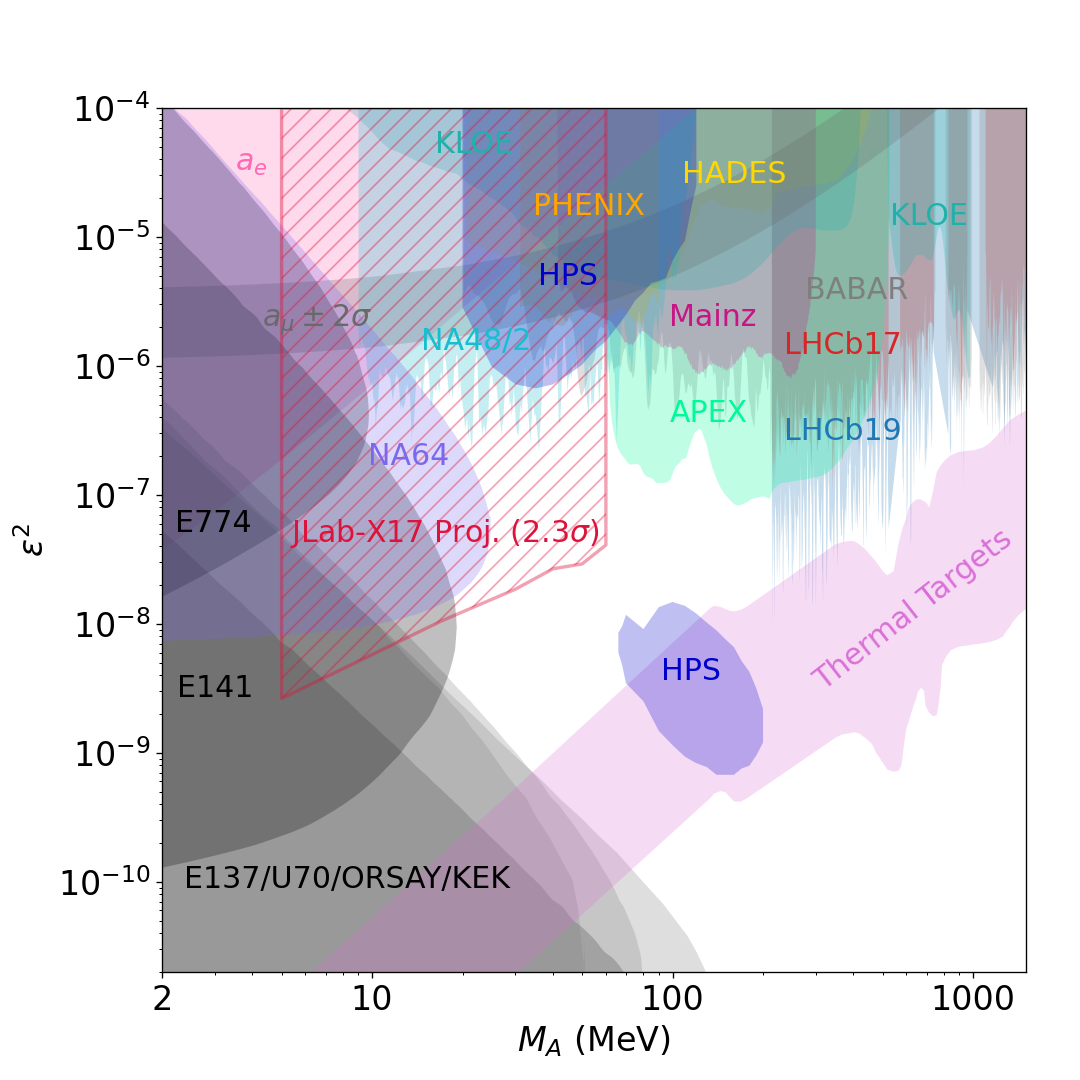}}
  \caption{Projected coverage of the square of the coupling constant ($\epsilon^2$) and mass ($m_{X}$) parameter space by this proposed experiment is shown as the thick red lines for the combined statistics of the two beam energies. The projections are superimposed on top of a constraints plot adapted from Ref.~\cite{Alexander:2016hsr}, with additional data from Refs.\cite{LHCb:2017trq,LHCb:2019vmc,Baltzell:2022rpd}.}
  \label{fig:proj_ps}
\end{figure}
In order to estimate the projected sensitivity to $\epsilon^2$, a comprehensive simulation of the experiment was carried out using the Geant4 simulation package developed for the PRad experiment~\cite{PRad:2019}. The simulation used a hybrid event generator that was built by combining the Madgraph5 generator~\cite{Conte:2012fm} for the Bethe-Heitler process  with the Geant4 generator (with Bethe-Heitler turned off) to utilize the best features of the two generators. The background was simulated for 200 s of beamtime and then scaled by sampling the number of events bin-by-bin to give the expected background for 30 days of 100 nA, 3.3 GeV beam.

%The projected sensitivity to 
%$\epsilon^2$ is dominated by the 3.3 GeV run due to lower %background level and better acceptance over the mass range %of search as seen in Fig.~\ref{fig:deteff}. 

The production rates for the $X$ particle are calculated using the rate equation (Eq. 14) from Ref.~\cite{Bjorken:2009mm} and rearranged to give the sentitivity as:
\begin{equation}
\label{eq:sensi}
\epsilon^2 = \frac{N_{X}}{{\mbox{5}} \times N_e T \frac{m_e^2}{m_X^2}},
\end{equation}
\noindent
where $N_X$ is the number of $X$ particles produced, $N_e$ is the number of incident electrons (1.62$\times$10$^{18}$ for 30 days at 100 nA), $T$ is the target thickness (2.5$\times$10$^{-4}$ r.l.), $m_X$ is the mass of the produced $X$, and $\epsilon^2$ is the square of the dimensionless coupling constant of the $X$ to SM matter. 

\noindent
The lowest $\epsilon^2$ achievable is calculated using 
Eq.~\ref{eq:sensi} with $N_X$ obtained by scaling 
the signal counts with the detection efficiency and impact of the decay length (shown in Fig.~\ref{fig:deteff}). 
The background event count is calculated by selecting events from the background simulation within a $\pm$~3$\sigma_{m_{X}}$ window. The typical invariant mass resolution $\sigma_{m_{X}}$ obtained from the simulation is shown in Fig.~\ref{fig:invmres}. The signal counts necessary given the background counts 
%listed in Table~\ref{tab:sensitivity_table}, 
are calculated using the criteria for a 2.3$\sigma$ significance used by other experiments (Eq.~\ref{eq:5sig}). 
%The mass resolutions are taken from the simulation of signal events shown in Fig.~\ref{fig:invmres}. 

\begin{equation}
\label{eq:5sig}
\frac{N_{\mbox{signal}}}{\sqrt{N_{\mbox{signal}} + N_{\mbox{bgd}}}} \geq {\mbox{2.3}}.
\end{equation}

\noindent
The 2.2 GeV runs will serve as reference runs to better understand the background and the signal, so it was conservatively estimated that 50\% of their statistics can be combined into the 3.3 GeV data for the final results. Using these range of sensitivities the bounds for the $\epsilon^2 - m_{X}$ parameter space is plotted in Fig.~\ref{fig:proj_ps} for the combined projected statistics.

\section{Conclusion}

The X17 anomaly suggests the existence of a new hidden sector particle.
It is of critical importance to resolve this anomaly quickly, as it not only could point to new physics but also has the potential to resolve additional anomalies.
In this white paper, we have described a nearly ready-to-run experiment at JLab using the PRad spectrometer that has the capability to fully address this anomaly.
By using a magnet-free spectrometer, two beam energies, and detecting all three final state the experiment will provide unprecedented control of the experimental background, leading to a results with uniquely low systematics.

\begin{acknowledgments}
This material is based upon work supported by the U.S. Department of Energy, Office of Science, Office of Nuclear Physics under contract DE-AC05-06OR23177.
\end{acknowledgments}

\bibliography{references}
\end{document}